\documentclass[%
 reprint,
 superscriptaddress,
 amsmath,
 amssymb,
 aps,
 prc,nofootinbib
]{revtex4-2}

\usepackage{hyperref}
\usepackage{amsmath}
\usepackage{amssymb}
\usepackage{graphicx}
\usepackage{mathrsfs}
\usepackage{color}
\usepackage{bm}
\usepackage{times}

\begin{document}

\title{Extension of the generator coordinate method with basis optimization}

\author{Moemi Matsumoto}
\affiliation{Department of Physics, Tohoku University, Sendai, 980-8578, Japan}
\author{Yusuke Tanimura}
\affiliation{Department of Physics, Tohoku University, Sendai, 980-8578, Japan}
\affiliation{Department of Physics and Origin of Matter and Evolution of Galaxy (OMEG) Institute, Soongsil University, Seoul 06978, Korea}
\author{Kouichi Hagino}
\affiliation{Department of Physics, Kyoto University, Kyoto, 606-8502, Japan}
\date{\today}

\begin{abstract}
   The generator coordinate method (GCM) has been a well-known method to describe nuclear collective motions.
   In this method, one specifies {\it a priori} the relevant collective degrees of freedom as input of the method, 
   based on empirical and/or phenomenological assumptions. 
   We here propose a new extension of the GCM, 
   in which both the basis Slater determinants and weight factors are optimized 
   according to the variational principle.
   Applying this method to $^{16}$O and $^{28}$Si nuclei with the Skyrme functional, 
   we demonstrate that the optimized bases correspond to excited states along a collective path, 
   unlike the conventional GCM which superposes only the local ground states. 
   This implies that a collective coordinate for large amplitude collective 
   motions is determined in a much more complex way than 
   what has been assumed so far. 
\end{abstract}


\maketitle

Collective motions are characteristic features of many-body systems. 
In atomic nuclei, giant resonances and low-lying surface vibrations with several multipolarities 
are typical examples. 
One of the unique features in nuclear systems 
is that large-amplitude collective motions often 
play essential roles in
phenomena such as nuclear fission and shape coexistence.
To describe such collective motions in a microscopic way is one of the major goals 
of quantum many-body problems.

In order to treat large-amplitude motions, the generator coordinate method (GCM) \cite{RingSchuck,Reinhard87}
has often been employed. 
The same GCM has been employed in describing alpha-clustering phenomena as well \cite{Kimura2012}. 
In GCM, one {\it a priori} specifies collective degrees of freedom (that is, collective coordinates), 
such as nuclear deformations and distances between clusters, 
and superposes many Slater determinants (SDs) within the selected collective subspace. 
The weight coefficients of SDs are then determined
according to the variational principle.
Many numerical calculations have been performed using this method, 
which is often referred to as the beyond-mean-field method \cite{Bender03,Niksic11,Robledo18,Yao10,Yao14}.

However, GCM has a serious problem that there always exists 
arbitrariness in the choice of collective coordinates, 
which one has to select in empirical and phenomenological manners.
With such choice, it is not trivial 
whether the collective motion of interest can be optimally described. 
In principle the collective coordinates can be extracted 
using the self-consistent collective coordinate (SCC) method \cite{Marumori80,Matsuo86,Matsuo00,Hinohara08}.
However, 
since the SCC does not consider multi-reference states, 
it is not obvious whether the collective coordinate determined by the SCC 
can be straightforwardly employed in the GCM. 
Furthermore, practical applications of dynamical GCM (DGCM) method, 
which introduces the conjugate momentum to collective coordinate, has been recently performed \cite{Hizawa21,Hizawa22}, 
pointing out 
that the conjugate momentum is an important degree of freedom in collective motions.
Therefore, a description of collective motions which avoids pre-set collective coordinates 
is always desirable
in order not to miss important degrees of freedom.

To discuss collective motion in a non-empirical way, 
several methods have been developed.
For example, a stochastic selection of the basis is carried out with anti-symmetrized molecular dynamics \cite{Itagaki03,Descouvemont18,Descouvemont20} 
and multi-reference density functional theory (DFT) \cite{Shinohara06,Fukuoka13}.
Even though the basis is constructed in an unbiased way with this method, 
it is not obvious whether such basis 
is optimal for the ground or low-lying excited states of a system.
Similar attempts have been taken 
in the Monte Carlo shell model (MCSM), in which 
the optimization of the basis 
is performed not only stochastically, but also variationally \cite{Shimizu12}.
Other examples of configuration-mixing methods with a basis optimization 
are the Multi-Configuration Self-Consistent Field (MCSCF)/the Multi-Configuration 
Hartree-Fock (MCHF) \cite{Schmidt98} in quantum chemistry.
A MCSCF/MCHF wave function is given by a superposition of particle-hole excitations  
on top of an optimized reference SD \cite{Faessler69}.
This is nothing but a configuration interaction (CI) method with an optimized 
mean-field potential. 
Even though the MCSCF/MCHF has been applied to several light nuclei
\cite{Faessler69,Satpathy70,Satpathy70-erratum,Pillet17,Robin16,Robin17}, 
the configurations generated by a single reference SD 
may not be efficient for describing nuclear large-amplitude motions.

In this Letter, we propose 
a new method without pre-fixed collective coordinates 
based on multi-reference DFT. 
In our new method, the trial many-body state is given by a 
superposition of SDs as in GCM, 
but both the weights and the basis SDs are determined by variation as in 
the MCSM \cite{Shimizu12}. 
A similar method has been considered in the time-dependent domain 
with various simplificiations \cite{Hasegawa20}. 
With this formulation, the collective space covered by the basis states is 
automatically optimized to the ground state of a system of interest.
We call this method the Optimized-basis GCM (OptGCM).

 Let us assume that the ground state wave function is given by 
 \begin{equation}
 |\Psi\rangle = \sum_{a=1}^M f_a|\Phi_a\rangle, 
 \label{eq:trial_gcm}
 \end{equation}
 where $|\Phi_a\rangle$ are the basis SDs  
 with the antisymmetrized product of $N$ orthonormal single-particle orbitals 
 $\varphi^{(a)}_i$($i=1,\ldots,N$). 
 %We use simplified notations, $|a_i\rangle$ for $|\varphi_i^{(a)}\rangle$. 
 We minimize the total energy for a given Hamiltonian $H$, 
  \begin{eqnarray}
 E = \frac{\langle\Psi|H|\Psi\rangle}{\langle\Psi|\Psi\rangle}
 =
 \frac{\sum_{ab}f_a^*f_b H_{ab}}{\sum_{ab}f_a^*f_b N_{ab}},
 \end{eqnarray}
 where $N_{ab}$ and $H_{ab}$ are the norm and Hamiltonian kernels, 
 \begin{eqnarray}
    N_{ab} &=& \langle\Phi_a|\Phi_b\rangle,\label{eq:normkernel} \\ 
    H_{ab} &=& \langle\Phi_a|H|\Phi_b\rangle, 
 \end{eqnarray}
respectively. 
The transition density matrix,
the energy density functional, and the Hartree-Fock (HF) Hamiltonian are defined as
\begin{eqnarray}
   \rho_{\beta\alpha}^{(ab)} &=& \frac{\langle\Phi_a|a_{\alpha}^\dagger a_{\beta}|\Phi_b\rangle}{\langle\Phi_a|\Phi_b\rangle},\\
   E^{(ab)}&=&E[\rho^{(ab)}]=H_{ab}/N_{ab},\\
   h_{\alpha\beta}^{(ab)}&=&\frac{\delta E[\rho^{(ab)}]}{\delta \rho_{\beta\alpha}^{(ab)}},
\end{eqnarray}
respectively, with the particle creation (annihilation) operator
$a^\dagger_\alpha$ ($a_\alpha$) of the single-particle state $\alpha$.
In the special case with $a=b$, the superscript $(ab)$ is denoted as $(a)$.
The total energy $E$ should be stationary under an arbitrary variation of the variational parameters
$f_a$ and $\varphi_i^{(a)}$. 
We take variation of $E$ with respect to $\langle \varphi_i^{(a)}|$ 
under the condition that the variation is orthogonal to the occupied 
orbitals in $|\Phi_a\rangle$, 
that is, the particle-hole type, 
which yields \cite{Shimizu12}
\begin{eqnarray}
   \frac{\delta E}{\delta\langle \varphi_i^{(a)}|_{ph}}&=&
   \sum_{b} 
   \frac{f_a^*N_{ab}f_b}{\langle\Psi|\Psi\rangle}(1-\rho^{(ab)})
    \nonumber
   \\
   &&
   \hspace{0.6cm}
   \times\left[
   E-E^{(ab)}+h^{(ab)}\rho^{(ab)}
   \right]|\varphi_i^{(a)}\rangle.
   \label{eq:dE/dphi}
\end{eqnarray}
Note that in the special case where there is only a single determinant,
this reduces to the gradient of the HF energy as given in Ref. \cite{Reinhard82}.
On the other hand, variation with respect to $f_a^*$ yields
\begin{eqnarray}
\frac{\delta E}{\delta f_a^*} &=&
\frac{1}{\langle\Psi|\Psi\rangle}\sum_b(H_{ab}-EN_{ab})f_b. 
\label{eq:dE/df}
\end{eqnarray}
Note 
that the ground state corresponds to 
$\frac{\delta E}{\delta\langle \varphi_i^{(a)}|_{ph}}=0$ and 
$\frac{\delta E}{\delta f_a^*}=0$. 
The latter condition 
leads to the Hill-Wheeler equation, 
which represents the generalized eigenvalue problem that arises 
also in the conventional GCM \cite{RingSchuck}. 

Starting from a set of initial conditions for the weights and basis SDs,
we search for the minimum of the total energy
by the conjugate gradient method \cite{NumericalRecipes}.
At each conjugate gradient step, the single-particle states within each SD may be  orthonormalized. 
In this work, the initial set of SDs are prepared by 
Woods-Saxon potentials with different deformations, and 
the initial values of the weight factors are set as $f_a=1$ for all $a$.
We have confirmed that the results are independent of the choice of the initial 
conditions for SDs and $f_a$. 
For the energy density functional, we adopt the SIII parameter set of 
the Skyrme functional \cite{Beiner75}. 
We omit the time-odd terms of the functional and the Coulomb interaction for simplicity. 
The pairing correlation is also neglected. 
The single-particle states are expanded on the 
axial harmonic-oscillator (HO) basis \cite{Vautherin73}. For simplicity, we restrict 
the weight factors $f_a$ to be real numbers. 

\begin{figure}
   \includegraphics[width=8.5cm]{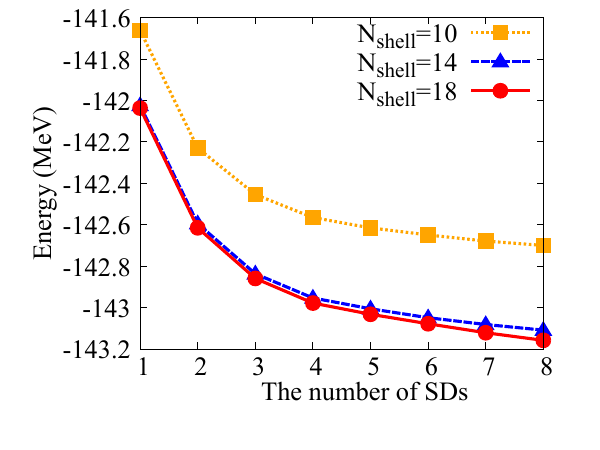}
   \caption{The ground state energy of $^{16}$O obtained with the OptGCM 
   as a function of the number of superposed SDs, $M$. 
   The calculations are performed for several truncations of the major harmonic oscillator 
   shells, $N_{\rm shell}$, for the basis expansions.
   }
   \label{fig:sd_energy}
\end{figure}

We now apply this method to the ground state of the $^{16}$O nucleus.
Figure \ref{fig:sd_energy} shows the ground state energy as a function of the number of SDs, $M$.
We calculate the energies for different numbers, 10, 14, and 18, of HO major shells 
for the basis set. 
For a given number of HO major shells, the ground state energy tends to 
converge as the number of SDs increases. 
It can also be seen that the ground state energy converges for 
the number of major shells as well.
In all the cases examined here, the beyond-mean-field correlations lower 
the ground state energies by about 1 MeV 
as compared to those of the HF approximation (the $M$ = 1 case).  

\begin{figure*}
   \includegraphics[width=18cm]{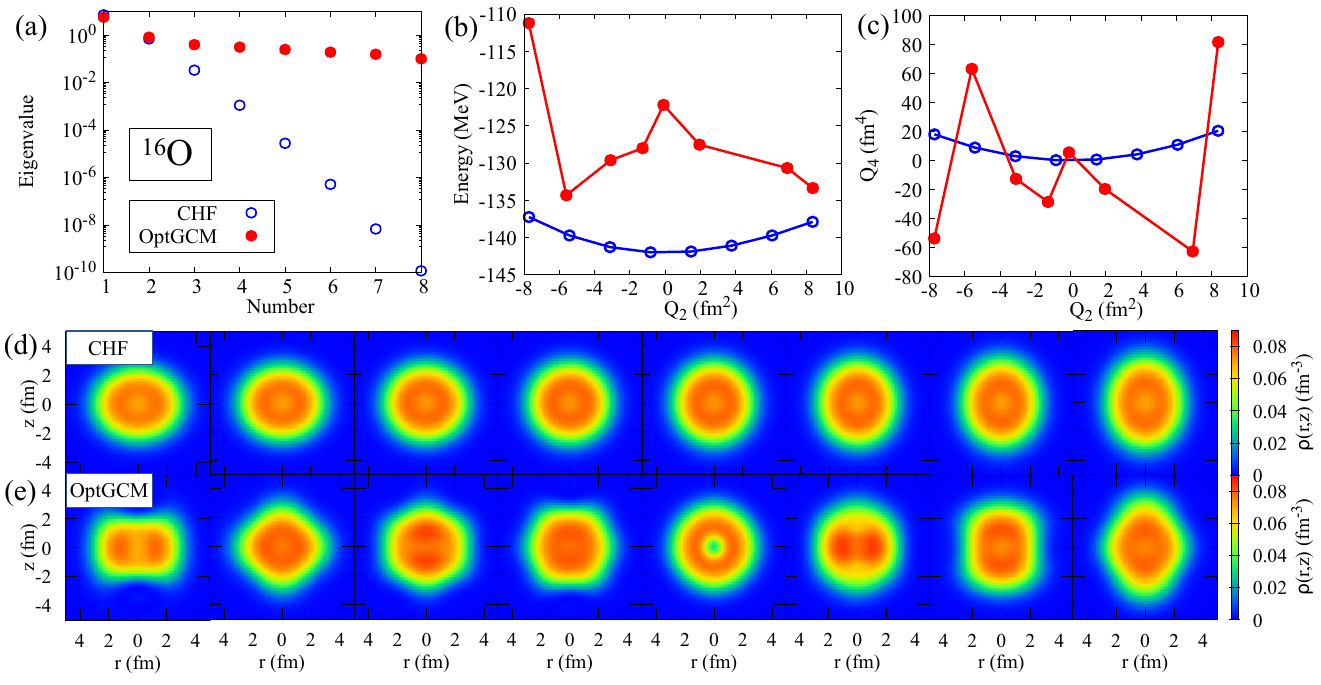}
   \caption{The properties of the basis SDs obtained with the constrained-HF (CHF) and 
   the OptGCM. 
   (a) The eigenvalues of the norm kernels plotted in the descending order. 
      The CHF and the OptGCM results are shown by the blue open circles and the red filled circles, respectively. 
   (b) The expectation value of the Hamiltonian as a function of the expectation value of $Q_2$ for each SD. The meaning of each symbol is the same as in the panel (a). 
   (c) The quadrupole ($Q_2$) and the hexadecupole ($Q_4$) moments of each SD in the two-dimensional $Q_2$-$Q_4$ plane. 
   The meaning of each symbol is the same as in the panel (a). 
   (d) The neutron density distributions of each SD obtained with the CHF. 
   (e) The same as the panel (d), but with the OptGCM.}
   \label{fig:o16_sd8}
\end{figure*}

Let us compare our results with those of the conventional GCM, 
in which the basis SDs are prefixed and only the weight factors $f_a$ are 
optimized by using Eq. \eqref{eq:dE/df}.
For this purpose, we take the results for 8 SDs with 14 HO major shells 
for the basis set.
Since the quadrupole collectivity is of primary importance in many nuclei, 
we take the quadrupole moment for the collective coordinates 
in our GCM calculation, {\it i.e.}, the GCM wave function is given as  
\begin{eqnarray}
   |\Psi_{\rm GCM}\rangle = \sum_{a=1}^{8}f_a|\Phi (Q_2^{(a)})\rangle, 
\end{eqnarray}
where the basis SDs $|\Phi(Q_2^{(a)})\rangle$ are 
the local ground state obtained with constrained Hartree-Fock (CHF) method 
with 
\begin{eqnarray}
   \langle\Phi_a(Q_2^{(a)})|\hat Q_2 | \Phi_a(Q_2^{(a)})\rangle = Q_2^{(a)}.
\end{eqnarray}
Here, 
\begin{eqnarray}
   \hat{Q}_\lambda&=&\int d^3r \ r^\lambda Y_{\lambda0}(\hat{\bm{r}})\hat \rho(\bm{r})
\end{eqnarray}
is the multipole moment operator with $\hat\rho$ and $Y_{\lambda0}$ 
being the density operator and the spherical harmonics, respectively.
For $Q_2^{(a)}$, we take the eight values equally spaced between $-$7.73 
fm$^2$ and 8.36 fm$^2$,
which are the smallest and largest values of the quadrupole moment of the optimized basis SDs
\footnote{We have confirmed that the results of the GCM calculation did not significantly 
altered even when the range of $Q_2^{(a)}$ was enlarged to the region $(-11,11)$ fm$^2$.}

In Table \ref{tab}, the ground state energies obtained with HF, GCM and OptGCM are summarized.
The GCM lowers the total energy by about 0.4 MeV relative to the HF.
On the other hand, in OptGCM,
the energy is further lowered by 0.7 MeV as compared to the GCM energy. 
This implies that the optimized basis states 
gain more correlations than the $Q_2$-CHF basis states. 

\begin{table}[]
   \caption{The ground state energies, the quadrupole moments, $Q_2$, 
   and the hexadecupole moments, $Q_4$, obtained with the 
   HF, GCM and OptGCM methods.}
   \label{tab}
   \begin{ruledtabular}
   \begin{tabular}{lccc}
   Method      & Energy (MeV)  &$Q_2$ (fm$^2$) &$Q_4$ (fm$^4$) \\ \hline
   HF          & $-142.024 $     & $0.000$ & $0.000$\\
   GCM         & $-142.407  $    & $0.179$ & $4.662$\\
   OptGCM & $-143.109 $    & $0.165 $ & $0.563$\\
   \end{tabular}
\end{ruledtabular}
\end{table}

To understand the difference 
between the OptGCM and the conventional GCM, 
Fig. \ref{fig:o16_sd8} analyzes the bases obtained with the OptGCM and 
the $Q_2$-CHF calculation.
Firstly, 
Fig. \ref{fig:o16_sd8} (a) shows the eigenvalues of the norm kernel, \eqref{eq:normkernel}.
In the CHF case, the eigenvalues range from $\sim 1$ to $10^{-10}$. 
The small eigenvalues of the norm kernel 
imply that the basis SDs are not fully linearly independent, which may result in 
the well-known overcompleteness problem. 
In contrast, all of the eigenvalues in the OptGCM case are larger than $\approx 10^{-1}$, 
which indicates that the optimized basis set effectively spans a larger subspace than 
the bases obtained with the CHF calculations.
In fact, we find that the GCM energy is not significantly altered even if we remove 
4 eigenstates of the norm kernel with the 4 lowest eigenvalues from the GCM calculation. 

Figure \ref{fig:o16_sd8} (b) shows the energies of each SD,
$E^{(a)}=E[\rho^{(a)}]=\langle \Phi_a|H|\Phi_a\rangle$, as a function of the 
quadrupole moment for each SD, $Q_2^{(a)}=\langle\Phi_a|\hat Q_2|\Phi_a\rangle$.
Interestingly, 
the optimized bases have higher energies than 
the energies with CHF. 
That is, the optimized bases correspond to 
excited states on top of the potential energy curve obtained with CHF.
In particular, there are some bases 
whose energies reach as high as 20 MeV or more above the potential energy curve. 
We note that a similar behavior is observed also in the case where 
the basis set is orthonormalized. 
Even if the expectation values, that is, the diagonal components, 
of the Hamiltonian $H$ with the OptGCM 
are larger than those with the GCM, the off-diagonal 
components with different SDs are larger either in the Hamiltonian or in the norm kernels, 
which result in the lower value of the ground state energy. 

Figure \ref{fig:o16_sd8} (c) shows the quadrupole moments $Q_2^{(a)}$ 
and the hexadecupole moment $Q_4^{(a)}=\langle\Phi_a|\hat Q_4|\Phi_a\rangle$ for each 
of the SDs on the two-dimensional $(Q_2,Q_4)$ plane. 
The filled and the open circles are for the SDs with the OptGCM and the $Q_2$-CHF, respectively. 
In the $Q_2$-CHF, the obtained states correspond to those along the valley of 
the two-dimensional potential energy surface, $E(Q_2,Q_4)$, and thus the variation of 
$Q_4$ is rather smooth. 
On the other hand, in the OptGCM, $Q_4$ fluctuates significantly around the valley and 
takes both positive and negative values.
We have found that the situation is similar for the $Q_6$ moment as well. 
As a result, the total $Q_4$ becomes smaller than the $Q_4$ in the $Q_2$-CHF. 
See Table \ref{tab} for the quadrupole and hexadecupole moments, 
\begin{eqnarray}
   Q_2 = \frac{\langle \Psi |\hat Q_2|\Psi\rangle}{\langle \Psi |\Psi\rangle}, \
   Q_4 = \frac{\langle \Psi |\hat Q_4|\Psi\rangle}{\langle \Psi |\Psi\rangle}
   % Q_4 = \int d^3r \ r^4 Y_{40}(\hat{\bm{r}})\rho(\bm{r})
\end{eqnarray}
of the total densities (see Table \ref{tab}).
One can see that while the values of $Q_2$ in the GCM and the OptGCM are 
almost the same, the value of $Q_4$ is smaller in the OptGCM.
Note that, although $Q_2$ and $Q_4$ have non-zero values for the both cases,
the total density distributions are nearly spherical.

The large fluctuation of $Q_4$ in the OptGCM indicates that 
the fluctuation of $Q_4$ as well as $Q_2$ is important in describing the ground state.
Moreover, Fig. \ref{fig:o16_sd8} (b) also indicates that considering only the local 
ground states is not sufficient. These two suggest that i) $Q_2$ alone does not make a good 
collective coordinate and at least one needs to take into account both $Q_2$ and $Q_4$, 
and ii) one needs to take into account excitations of nuclei in determining a collective 
coordinate. Notice that 
a recent GCM calculation includes not only $Q_2$ but also $Q_4$ as collective coordinates \cite{Kumar23}. Such calculation takes into account the point i) above, but 
one would need to include not only the local ground states but also excited states 
to take into account the point ii). 
While such calculation with conventional GCM with the prefixed collective subspace 
would be practically and numerically demanding, 
both the effects are automatically taken into account with the OptGCM. 
This advantage allows one not only to find out appropriate collective coordinates 
but also to avoid technical problems and numerical efforts of performing a 
multi-dimensional GCM.

Figures \ref{fig:o16_sd8} (d) and \ref{fig:o16_sd8} (e) show  
the neutron density distributions of each SD obtained with the $Q_2$-CHF and the OptGCM, 
respectively. 
With the OptGCM, a variety of shapes are obtained for the bases, that cannot be 
obtained with the $Q_2$-CHF. 
In particular, we find a bubble-like basis and a $Q_6$ deformed basis.
This suggests that the OptGCM is capable of automatically describing multiple 
collective modes. 

\begin{figure}[bt]
   \includegraphics[width=7.1cm]{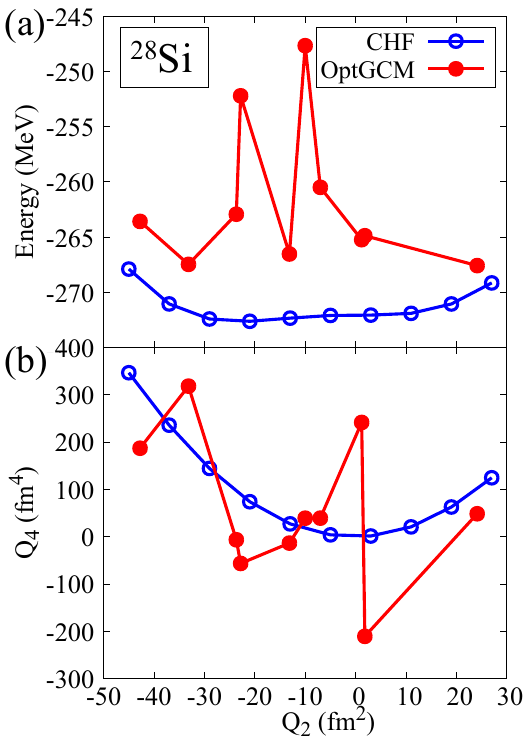}
   \caption{Same as Figs. \ref{fig:o16_sd8}(b) and \ref{fig:o16_sd8}(c), but for the  
   $^{28}$Si nucleus. 
   }
   \label{fig:28Si_sd10}
\end{figure}

We next show the results for the $^{28}$Si nucleus, 
which has an oblate ground state in the HF approximation with a soft energy curve 
in the $Q_2$ direction. 
The ground state energies are obtained to be  
$-272.63$, $-273.95$, and $-275.00$ MeV with the HF, the GCM, and the OptGCM, respectively. 
As in $^{16}$O, a further decrease of the ground state energy is achieved with the OptGCM, 
for which the energy difference between the GCM and the OptGCM is 1.05 MeV. 
That is, the OptGCM gains more correlation energy than the GCM.  
Figs. \ref{fig:28Si_sd10}(a) and \ref{fig:28Si_sd10}(b) show the energies 
and the multipole moments of each 
SD, respectively. As in the case of $^{16}$O shown in Fig. \ref{fig:o16_sd8}, 
the bases with the OptGCM correspond to excited states and the $Q_4$ are largely 
fluctuated. From this analysis, it is evident that the conventional GCM with local 
ground states is not sufficient to describe the ground state of atomic nuclei. 

In summary, 
we have proposed a novel extension of the GCM 
in which both the basis SDs 
and the weight factors are optimized according to the variational principle.
We have applied this method to $^{16}$O and $^{28}$Si nuclei with the Skyrme energy functional,
and have found that the optimized SDs correspond to excited states rather than the local 
ground states. This leads to an important question concerning to what extent the conventional 
GCM, which superposes only local ground states along a pre-set collective coordinate, 
is justified. In particular, we have found that the quadrupole moment $Q_2$ alone may not make 
an appropriate collective coordinate for the ground state of $^{16}$O and $^{28}$Si, 
and at least the hexadecapole moment $Q_4$ also has to be taken into account, even though 
$Q_2$ and $Q_4$ would not exhaust the relevant collective space. 

It would be an interesting future work to apply the present method 
systematically to many nuclei in the nuclear chart and 
to compare with other existing methods for collective motions, 
such as the conventional GCM, the SCC method 
\cite{Marumori80,Matsuo86,Matsuo00,Hinohara08}, 
and the dynamical GCM with collective momenta \cite{Hizawa21,Hizawa22}. 
Such comparison would 
identify 
the collective coordinates relevant to nuclear collective phenomena. 
Another interesting future direction is to apply this method to excited states and 
discuss a spectrum of a nucleus. 
To this end, 
one would need to investigate how basis SDs are optimized 
for the ground state and excited states. 
For this purpose, one would also need to implement 
the angular momentum projection 
to restore the broken rotational symmetry. 
We leave these for a future study. 

{\it Acknowledgments.}
We thank T. Abe, G. Col\`o, 
N. Hinohara, D. Lacriox, H. Sagawa, and K. Washiyama for useful discussions.
This work was supported by JST SPRING, Grant Number JPMJSP2114 and the JSPS KAKENHI Grant No. 19K03861. 
M. M. acknowledges the support from Graduate Program on Physics for
the Universe (GP-PU) of Tohoku University.
The numerical calculations were performed with the computer facility at the Yukawa Institute for Theoretical Physics, Kyoto University.
\bibliography{multisd}

\end{document}